\documentclass[
twocolumn,
superscriptaddress,
amsmath,amssymb,
aps,
]{revtex4-1}

\usepackage{bm}
	\usepackage{braket}
	\usepackage{dsfont}
 	\usepackage{graphicx}
	\usepackage[caption=false]{subfig}
	\usepackage[export]{adjustbox}
	\usepackage{amsmath}
	\usepackage{amssymb}
	\usepackage{hyperref}		% hyperlinks in the pdf, has to be (almost) last %
	\usepackage[figure]{hypcap}		% corrects hyperlinks to figures %
	\usepackage{upgreek}
	\hypersetup{
	   	bookmarks		= false,		% show bookmarks bar %
		unicode			= false,	% non-Latin characters in Acrobat's bookmarks
		pdfborder			= {0 0 0}	% style of the border around a link; {0 0 0} gives no border %
		pdftoolbar			= false,		% show Acrobat's toolbar %
		pdfmenubar		= true,		% show Acrobat's menu %
		pdffitwindow		= false,     	% window fit to page when opened %
		pdfstartview		= {FitH},    	% fits the width of the page to the window %
		pdfnewwindow	= true,      	% links in new window %
		colorlinks			= true,		% false: boxed links; true: colored links %
		linkcolor			= blue,		% color of internal links %
		citecolor			= blue,		% color of links to bibliography %
		filecolor			= blue,		% color of file links %
		urlcolor			= blue,		% color of external links %
		linktocpage		= true,		% whole entry or only page in the toc is made into a link %
		pdftitle				= {Determination of interatomic coupling between two-dimensional crystals using angle-resolved photoemission spectroscopy},
		pdfauthor			= {M. Mucha-Kruczynski},
		pdfsubject			= {},
		pdfcreator			= {MikTeX},
		pdfproducer		= {Marcin Mucha-Kruczynski},
		pdfkeywords		= {},
		}

	\newcommand{\vect}[1]{\boldsymbol{\mathrm{#1}}}		% I use command \vect{} for vectors
	\newcommand{\op}[1]{\hat{#1}}	% I use command \op{} for operators

\begin{document}
\title{Determination of interatomic coupling between two-dimensional crystals using angle-resolved photoemission spectroscopy}

\author{J.~J.~P.~Thompson}
\affiliation{Department of Physics, University of Bath, Claverton Down, Bath BA2 7AY, United Kingdom}
\affiliation{Chalmers University of Technology, Department of Physics, SE-412 96 Gothenburg, Sweden}
\author{D.~Pei}
\affiliation{Clarendon Laboratory, Department of Physics, University of Oxford, Oxford OX1 3PU, UK}
\author{H.~Peng}
\affiliation{Clarendon Laboratory, Department of Physics, University of Oxford, Oxford OX1 3PU, UK}
\author{H.~Wang}
\affiliation{Center for Nanochemistry, Beijing National Laboratory for Molecular Sciences, Peking University, Beijing 100871, P.~R.~China}
\author{N.~Channa}
\affiliation{Department of Physics, University of Bath, Claverton Down, Bath BA2 7AY, United Kingdom}
\affiliation{Department of Physics, University of Warwick, Coventry CV4 7AL, United Kingdom}
\author{H.~L.~Peng}
\affiliation{Center for Nanochemistry, Beijing National Laboratory for Molecular Sciences, Peking University, Beijing 100871, P.~R.~China}
\author{A.~Barinov}
\affiliation{Elettra-Sincrotrone Trieste ScPA, Trieste 34149, Italy}
\author{N.~B.~M.~Schr\"{o}ter}
\affiliation{Clarendon Laboratory, Department of Physics, University of Oxford, Oxford OX1 3PU, UK}
\affiliation{Swiss Light Source, Paul Scherrer Institute, 5232 Villigen, Switzerland}
\author{Y.~Chen}
\affiliation{Clarendon Laboratory, Department of Physics, University of Oxford, Oxford OX1 3PU, UK}
\author{M.~Mucha-Kruczy\'{n}ski}
\email{M.Mucha-Kruczynski@bath.ac.uk}
\affiliation{Department of Physics, University of Bath, Claverton Down, Bath BA2 7AY, United Kingdom}
\affiliation{Centre for Nanoscience and Nanotechnology, University of Bath, Claverton Down, Bath BA2 7AY, United Kingdom}

\begin{abstract}
Lack of directional bonding between two-dimensional crystals like graphene or monolayer transition metal dichalcogenides provides unusual freedom in selection of components for vertical van der Waals heterostructures. However, even for identical layers, their stacking, in particular the relative angle between their crystallographic directions, modifies properties of the structure. We demonstrate that the interatomic coupling between two two-dimensional crystals can be determined from angle-resolved photoemission spectra of a trilayer structure with one aligned and one twisted interface. Each of the interfaces provides complementary information and together they enable self-consistent determination of the coupling. We parametrize interatomic coupling for carbon atoms by studying twisted trilayer graphene and show that the result can be applied to structures with different twists and number of layers. Our approach demonstrates how to extract fundamental information about interlayer coupling in a stack of two-dimensional crystals and can be applied to many other van der Waals interfaces.
\end{abstract}

\maketitle
\section{Introduction} 
Following the isolation of graphene (a layer of carbon atoms arranged in regular hexagons) in 2004 \cite{novoselov_science_2004}, many other atomically thin two-dimensional crystals have been produced and can be stacked in a desired order on top of each other. In contrast to conventional heterostructures, in which chemical bonding at interfaces between two materials modifies their properties and requires lattice matching for stability, stacks of two-dimensional crystals are held together by weak forces without directional bonding. As a result, any two of these materials can be placed on top of each other, providing extraordinary design flexibility  \cite{geim_nature_2013, novoselov_science_2016, liu_natrevmat_2016}. Moreover, subtle changes in atomic stacking, especially the angle between the crystallographic axes of two adjacent layers, can have big impact on the properties of the whole heterostructure, with examples including the observation of Hofstadter's butterfly \cite{ponomarenko_nature_2013, dean_nature_2013} and interfacial polarons \cite{chen_nanolett_2018} in graphene/hexagonal boron nitride heterostructures, interlayer excitons in transition metal dichalcogenide bilayers \cite{fang_pnas_2014, rivera_natnano_2018}, appearance of superconductivity in magic-angle twisted bilayer graphene \cite{cao_nature_2018, cao_nature_2018_2} and explicit twist-dependence of transport measurements in rotatable heterostructures \cite{chari_nanolett_2016, ribeiro-palau_science_2018, finney_natnano_2019}. Phenomena like these arise because the misalignment of two crystals changes the atomic registry at the interface and hence tunes the spatial modulation of interlayer interaction. Consequently, understanding the coupling between two two-dimensional materials at a microscopic level is crucial for efficient design of van der Waals heterostructures.

The impacts of a twisted interface and modulated interlayer coupling on the electronic properties of two-dimensional crystals include band hybridization \cite{ohta_prl_2012, diaz_nanolett_2015, wilson_sciadv_2017}, band replicas and minigaps due to scattering on moir\'{e} potential \cite{ohta_prl_2012, pierucci_nanolett_2016, ulstrup_sciadv_2020}, charge transfer and vertical shifting of bands \cite{yeh_nanolett_2016, zribi_npj2D_2019, wilson_sciadv_2017} as well as changes of the effective masses \cite{yeh_nanolett_2016, wilson_sciadv_2017}. Variations in the interlayer coupling as a function of the twist angle, $\theta$, were probed for example using photoluminescence, Raman and angle-resolved photoemission (ARPES) spectroscopies \cite{zande_nanolett_2014, huang_nanolett_2014, liu_naturecomms_2014, yeh_nanolett_2016}. Here, we use the last of those methods to image directly the electronic bands in trilayer graphene with one perfect and one twisted interface. From our data, we extract the interatomic coupling, $t(\vect{r}, z)$, describing coupling between two carbon atoms separated by a vector $\vect{r}_\text{3D} = (\vect{r},z) = (x,y,z)$. Such coupling functions, usually based on comparisons to ab initio calculations, can be used to determine electron hoppings in tight-binding \cite{laissardiere_nanolett_2010, fang_prb_2015} and continuum \cite{santos_prl_2007, wallbank_prb_2013} models of corresponding van der Waals interfaces at any twist angle. We show that $t(\vect{r}, z)$ determined purely by measurements on one of the structures accurately describes electronic dispersions obtained for stacks with different $\theta$ and number of layers, providing an experimentally verified set of parameters to model twistronic graphene. Our approach makes use of the fact that a trilayer structure is the thinnest stack that can contain both a perfect and twisted interface. The former, due to translational symmetry, can be straightforwardly described in the real space using $t(\vect{r}, z)$. At the same time, the impact of the moir\'{e} pattern formed at the latter can be captured in the reciprocal space by considering scattering by moir\'{e} reciprocal vectors on the momentum-dependent potential $\tilde{t}(\vect{q}, z)$ which is a two-dimensional Fourier transform $\mathcal{F}[t(\vect{r}, z)]$ of $t(\vect{r}, z)$ (see the comparison of the two cases in Fig.~\ref{fig:lattice}(a)). As a consequence, this method should enable determination of interatomic couplings for all van der Waals interfaces for which moir\'{e} effects were observed.

\section{Results}

\begin{figure}[t]
	     \centering
	     \includegraphics[width=1.00\columnwidth]{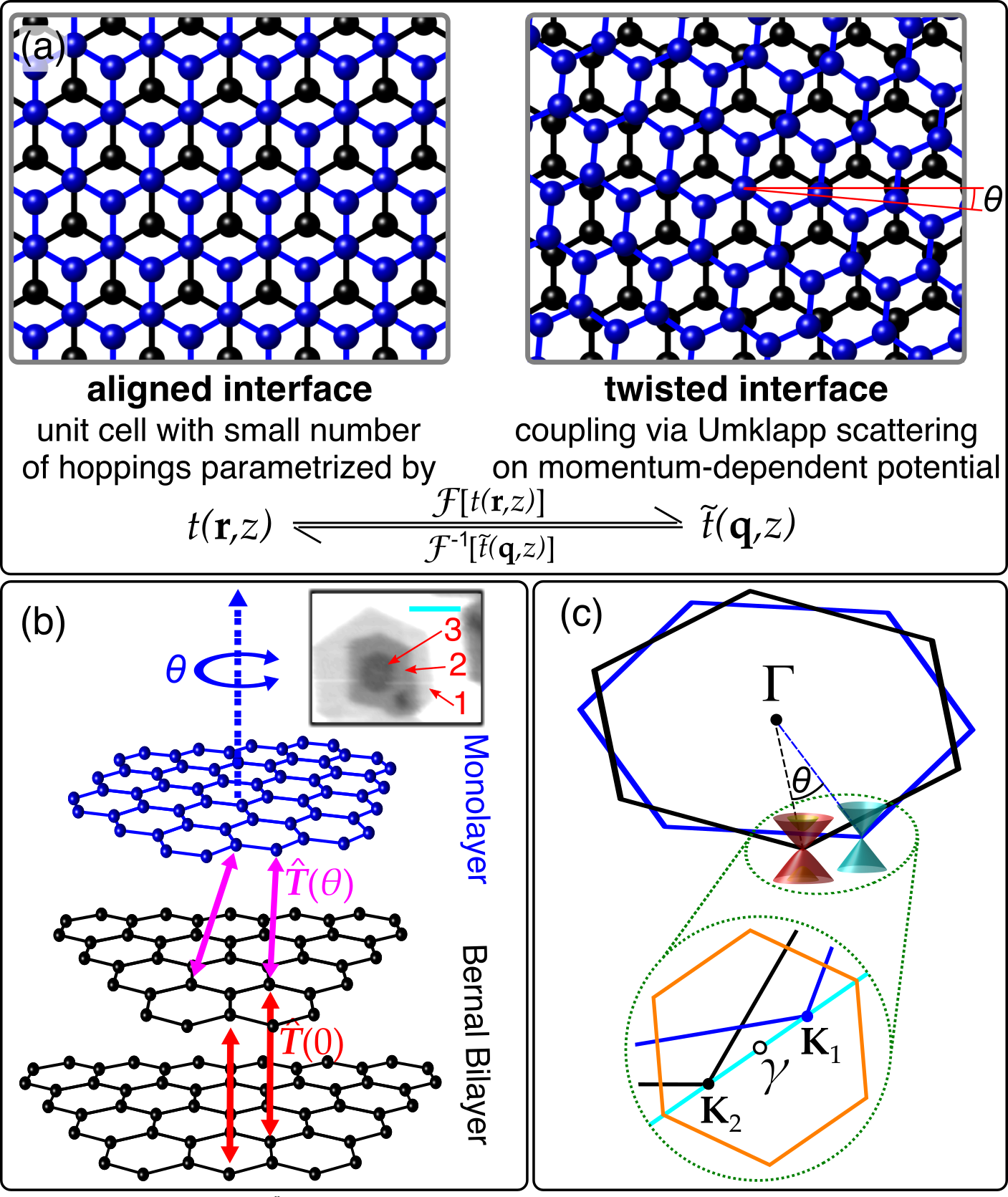}
	     \caption{\textbf{Aligned vs twisted interfaces in van der Waals heterostructures.} (a) Comparison of aligned and twisted interfaces for two-dimensional crystals and the descriptions in the real and reciprocal spaces used in this article. Blue and black balls indicate atoms in the top and bottom layer, respectively. (b) Schematic of twisted trilayer graphene with monolayer (blue) stacked at an angle on top of a Bernal bilayer (black). The red and purple arrows indicate the interlayer couplings for the Bernal and twisted interfaces which are captured by the blocks $\op{T}(0)$ and $\op{T}(\theta)$, respectively, in the Hamiltonian in Equation \eqref{ttlg}. Inset shows photoemission intensity from copper substrate which is attenuated by graphene layers above, providing a measure of graphene layer number. The red arrows indicate each of the graphene layers in the trilayer stack and the cyan line corresponds to the distance of 10 $\upmu$m. (c) Brillouin zones of the Bernal bilayer (black) and rotated monolayer (blue) with bilayer and monolayer graphene low-energy electronic spectra shown in the vicinities of one set of the Brillouin zone corners. The inset depicts in orange the superlattice Brillouin zone and the cyan line indicates the $k$-space path cuts along which are presented in Fig.~\ref{fig:arpes}(a) and \ref{fig:906}.}\label{fig:lattice}
\end{figure}

{\bf ARPES of twisted trilayer graphene} We grew our graphene trilayers on copper foil using chemical vapour deposition \cite{peng_advmat_2017, mattevi_jmc_2011}. The inset of Fig.~\ref{fig:lattice}(b) shows the intensity map of copper $d$-band photoelectrons which are attenuated differently by the overlying graphene layers depending on their number. This provides means to identify all of the layers in our stack, shown in the inset with different shades of gray and indicated with the red arrows. As depicted schematically in the main panel of Fig.~\ref{fig:lattice}(b), the bottom two layers form a Bernal bilayer (2L) while the crystallographic axes of the top monolayer (1L) are rotated by an angle $\theta$ with respect to those of the layer underneath. As a result, the Brillouin zones corresponding to the bilayer and monolayer are also rotated with respect to each other, Fig.~\ref{fig:lattice}(c). We focus here on the vicinity of one set of the corners of the two Brillouin zones, which we denote $\vect{K}_{\mathrm{2}}$ and $\vect{K}_{\mathrm{1}}$, for the bilayer and monolayer, respectively. The separation between these two points, dependent on the twist angle, defines an effective superlattice Brillouin zone, indicated in orange in the inset of Fig.~\ref{fig:lattice}(c).

\begin{figure}[t]
	     \centering
	     \includegraphics[width=1.00\columnwidth]{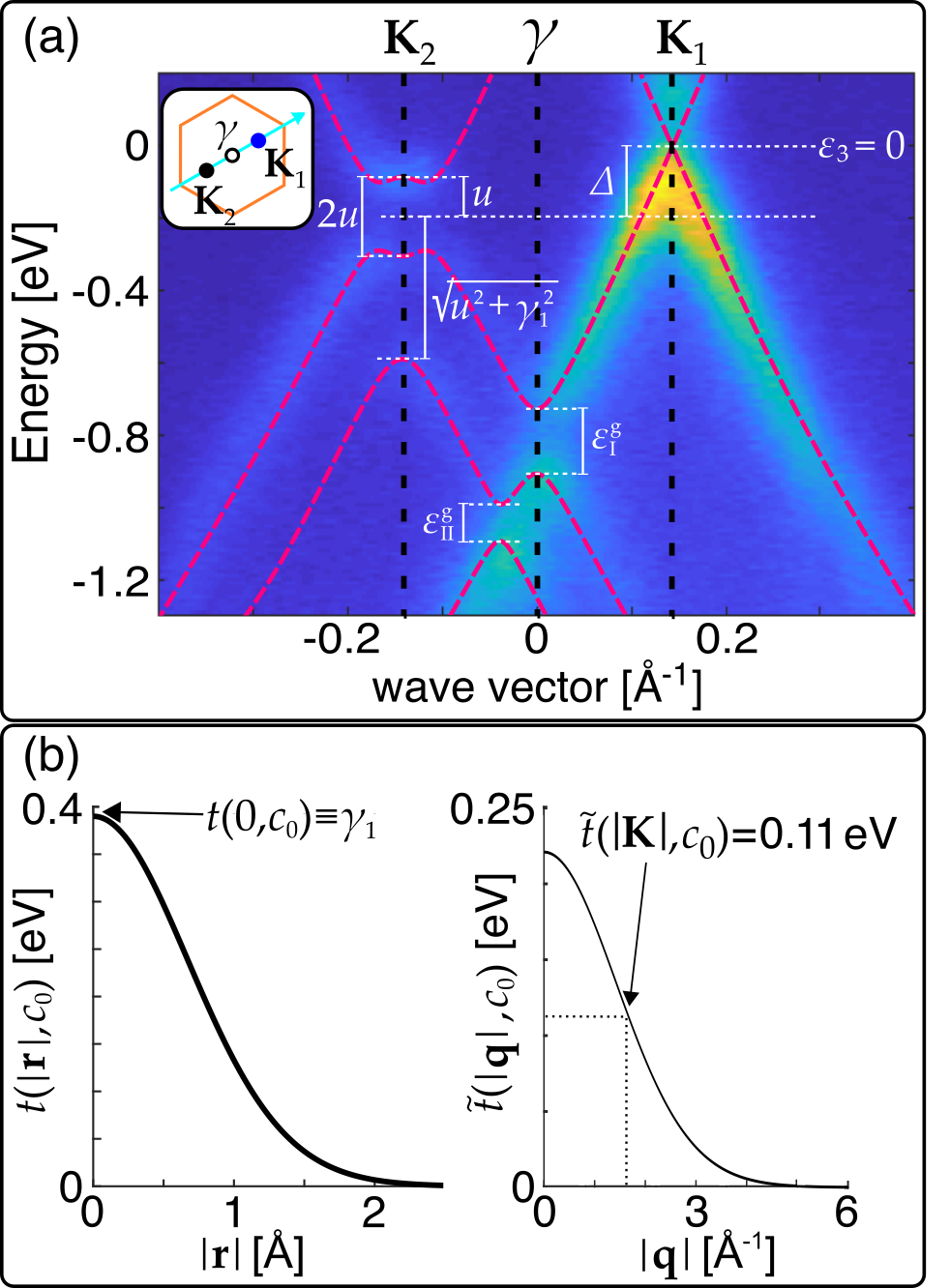}
	     \caption{\textbf{Angle-resolved photoemission spectra and interatomic coupling.} (a) ARPES intensity for twisted trilayer with twist $\theta=9.6^\circ$, measured along the direction connecting Brillouin zone corners $\vect{K}_{2}$ and $\vect{K}_{1}$ as shown in Fig.~\ref{fig:lattice}(c) and indicated in the inset. The calculated miniband structure along the same path is shown with red dashed lines. White dashed and solid lines indicate the important energies used to fit the parameters of our theoretical model. (b) Left: Real-space interatomic coupling, $t(|\vect{r}|,c_0)$, as a function of distance $|\vect{r}|$ between carbon atoms, as given in Equation \eqref{ansatz} with parameter values from Table \ref{table}. Right: Two-dimensional Fourier transform $\tilde{t}(|\vect{q}|,c_0)$ of the interatomic coupling, $t(|\vect{r}|,c_0)$, as a function of wave vector $|\vect{q}|$.}\label{fig:arpes}
\end{figure}

In Fig.~\ref{fig:arpes}(a), we present ARPES intensity along a cut in the $k$-space connecting $\vect{K}_{\mathrm{2}}$ and $\vect{K}_{\mathrm{1}}$, with the energy reference point set to the linear crossing (Dirac point) at $\vect{K}_{\mathrm{1}}$. Close to each corner, the intensity reflects the low-energy band structures of unperturbed 2L and 1L. Because the bilayer flake is below the monolayer, signal from the former is attenuated due to the electron escape depth effect. In between the two spectra, coupling of the two crystals leads to anticrossings of the bands and opening of minigaps (marked as $\varepsilon^{\mathrm{g}}_{\mathrm{I}}$ and $\varepsilon^{\mathrm{g}}_{\mathrm{II}}$ in the figure). As the size of the superlattice Brillouin zone depends on the twist angle, the energy positions of the minigaps also depend on $\theta$. Moreover, the magnitudes of the minigaps depend on the interlayer coupling between the bilayer and monolayer and also, in principle, vary with $\theta$. However, fundamentally, all of the features in our spectrum originate in interactions between carbon atoms, be it in the same or different layers, at the twisted or aligned interface. This provides us with an opportunity to study the interatomic coupling $t(\vect{r}, z)$ in carbon materials. 

{\bf Parametrizing carbon-carbon interaction potential} In order to understand our data, we use a generic Hamiltonian for a van der Waals heterostructure comprised of three layers of the same two-dimensional crystal
\begin{align} \label{ttlg}
\op{H}= \begin{pmatrix}
      \op{H}_0\left(0, \Delta-u \right) & \op{T}(0) & 0 \\
      \op{T}^\dagger(0)& \op{H}_0\left(0,\Delta +u \right)&\op{T}(\theta) \\
      0& \op{T}^\dagger(\theta) & \op{H}_0\left(\theta,0\right)
      \end{pmatrix}.
\end{align}
In this Hamiltonian, the diagonal block, $\op{H}_0\left(\theta_i, \varepsilon_i\right)$ describes the  $i$-th layer at a twist angle $\theta_i$, with on-site energies of atomic sites in this layer, $\varepsilon_i$. Here, because only the relative twist between any two adjacent layers is important, we have $\theta_1 = \theta_2 = 0$ and $\theta_3 = \theta$. Also, our choice of energy reference point is equivalent to $\varepsilon_{3}=0$ and we introduce potential energy difference, $2u=\varepsilon_1-\varepsilon_2$, as well as average energy, $\Delta =(\varepsilon_1+\varepsilon_2)/2$, of layers 1 and 2 (the charge transfer between the copper foil and the graphene layers giving rise to $u\neq\Delta\neq0$ is discussed in more detail in Ref.~\onlinecite{peng_advmat_2017}).  For graphene, the intralayer blocks $\op{H}_0$ can be straight-forwardly described using a tight-binding model  \cite{castro_neto_rmp_2009} for a triangular lattice with two inequivalent atomic sites, $A$ and $B$, per unit cell and nearest neighbour coupling between them $\gamma_0\equiv-t(\vect{r}_{AB},0)$, where $\vect{r}_{AB}$ is a vector connecting neighbouring $A$ and $B$ atoms with the carbon-carbon bond length $|\vect{r}_{AB}| = 1.46$ \AA.

Of more importance for us, however, are the off-diagonal blocks $\op{T}(\theta_i-\theta_{i-1})$ which capture the twist-dependent interlayer interactions between adjacent layers (we neglect the interaction between the bottom and the top layers which is at least an order of magnitude weaker \cite{koshino_prb_2009}). As the bottom two layers are stacked according to the Bernal stacking, a real-space description of the interlayer interaction block $\op{T}(0)$ is possible with the leading coupling $t(0,c_0) \equiv \gamma_1$, with interlayer distance $c_0 = 3.35$ \AA, due to atoms with neighbours directly above or below them, as shown in Fig.~\ref{fig:lattice}(a) \cite{mccann_prl_2006}. In contrast, we describe the coupling between the twisted layers, $i=2,3$, in the reciprocal space based on electron tunnelling from a state with wave vector $\vect{k}$ in layer 2 to a state with wave vector $\vect{k'}$ in layer 3 with the requirement that crystal momentum is conserved \cite{bistritzer_pnas_2011, koshino_njp_2015}, $\vect{k} + \vect{G} = \vect{k'} +\vect{G'}$, where $\vect{G}$ and $\vect{G'}$ are the reciprocal vectors of layers 2 and 3, respectively. The strength of a given tunnelling process is set by the two-dimensional Fourier transform, $\mathcal{F}[t(\vect{r},z)] = \tilde{t}(\vect{q},z)$, of the real-space coupling $t(\vect{r},z)$ so that
\begin{align} \label{intercoup}
 \op{T}(\theta) = &\sum_{\vect{G}, \vect{G'}}\tilde{t}(\vect{k}+\vect{G},z) \\
 &\times \begin{pmatrix}
 \mathrm{e}^{\mathrm{i} \vect{G} \cdot \vect{\tau}} &\mathrm{e}^{\mathrm{i} (\vect{G}+\op{R}_{\theta}\vect{G'}) \cdot \vect{\tau} } \\ \notag
 1 & \mathrm{e}^{\mathrm{i} \op{R}_{\theta}\vect{G'} \cdot \vect{\tau}}
 \end{pmatrix} \delta_{\vect{k}+\vect{G},\vect{k'}+\vect{G'}},
\end{align}
where $\vect{\tau} = (-|\vect{r}_{AB}|,\, 0)$ and $\op{R}_\theta$ is a matrix of clockwise rotation by angle $\theta$ (see Supplementary Note 1 for more details on the construction of the Hamiltonian $\op{H}$). 

The uniqueness of a trilayer with one perfect and one twisted interface (as exemplified in Fig.~\ref{fig:lattice}(a) for the case of graphene) lies in the fact that the Hamiltonian $\op{H}$ contains interlayer blocks based on both the real-space ($\op{T}(0)$) and reciprocal-space ($\op{T}(\theta)$) descriptions which provide complementary information and at the same time are related to each other because of the Fourier transform connection between $t(\vect{r},z)$ and $\tilde{t}(\vect{q},z)$. Because of this, comparison of the photoemission data with the spectrum calculated based on Equation \eqref{ttlg} provides more information about the interatomic coupling $t(\vect{r},z)$ than structures with one type of interface only. For our graphene trilayer, we compute the miniband spectrum of $\op{H}$ (see Methods for more details) assuming a Slater-Koster-like two-centre ansatz for $t(\vect{r},z)$ \cite{laissardiere_nanolett_2010},
\begin{align}\label{ansatz} 
t(\vect{r},z) & = t(|\vect{r}|,z) \\ &= V_{\uppi}(\vect{r},z) \left(1- \dfrac{z^2}{|\vect{r}_\text{3D}|^2}\right) + V_{\upsigma}(\vect{r},z)  \left(\dfrac{z}{|\vect{r}_\text{3D}|} \right)^2, \nonumber\\
V_{\uppi}(\vect{r},z) &= -\gamma_0 \exp\left[- \alpha_{\uppi}(|\vect{r}_\text{3D}|-|\vect{r}_{AB}|)\right],\nonumber\\
V_{\upsigma}(\vect{r},z) &= \gamma_1 \exp\left[-\alpha_\upsigma(|\vect{r}_\text{3D}|-c_0)\right],\nonumber
\end{align}
where $V_{\uppi}$ and $V_\upsigma$ represent the strength of the $\uppi$ and $\upsigma$ bonding \cite{koster_physrev_1954}, respectively, and $\alpha_\uppi$ and $\alpha_\upsigma$ their decay with increasing interatomic distance. 

In fitting our numerical results to the experimental data in Fig.~\ref{fig:arpes}(a), we first determine the position of 1L Dirac point what sets the $\varepsilon=0$ reference point. We then use the electronic band gap at $\vect{K}_{\mathrm{2}}$ to fix the electrostatic potential $2u$ and position the bilayer neutrality point halfway in the gap, establishing the potential energy shift $\Delta$. We obtain the in-plane nearest neighbour hopping $\gamma_{0}$ from the slope of the 1L linear dispersion close to the Dirac point at $\vect{K}_{\mathrm{1}}$ while the direct interlayer coupling $\gamma_{1}$ is set by the splitting of the 2L lower valence band from the neutrality point at $\vect{K}_{\mathrm{2}}$. Finally, the decay constants $\alpha_{\uppi}$ and $\alpha_{\upsigma}$ are found numerically using the constraints that (i) the magnitudes of the gaps $\varepsilon^{\mathrm{g}}_{\mathrm{I}}$ and $\varepsilon^{\mathrm{g}}_{\mathrm{II}}$ in Fig.~\ref{fig:arpes}(a) match the experimental data and (ii) in the limit of $\theta=0$, $\op{T}(\theta)$ from Equation \eqref{intercoup} converges to the real-space form of $\op{T}(0)$ as used for coupling between the Bernal stacked layers (see Supplementary Note 2 for further discussion).

\begin{table}[b] 
\begin{center}
\begin{tabular}{|c|c|c|c| }
\hline
\multicolumn{4}{|c|}{Fitting constants} \\
\hline
$\gamma_0$ [eV] & $\gamma_1$ [eV] & $\alpha_\uppi$ [$\AA^{-1}$] & $\alpha_\upsigma$ [$\AA^{-1}$]\\
\hline
2.95& 0.39 &3.39 & 6.78\\
\hline
\end{tabular}
\end{center}
\caption {\textbf{Parametrization of $t(|\vect{r}|,z)$} Values of fitting parameters describing the carbon-carbon potential $t(|\vect{r}|,z)$ from Equation \eqref{ansatz}.}\label{table}%\label{Table} 
\end{table}

\begin{figure}[t]
	\centering
	\includegraphics[width=1.00\columnwidth]{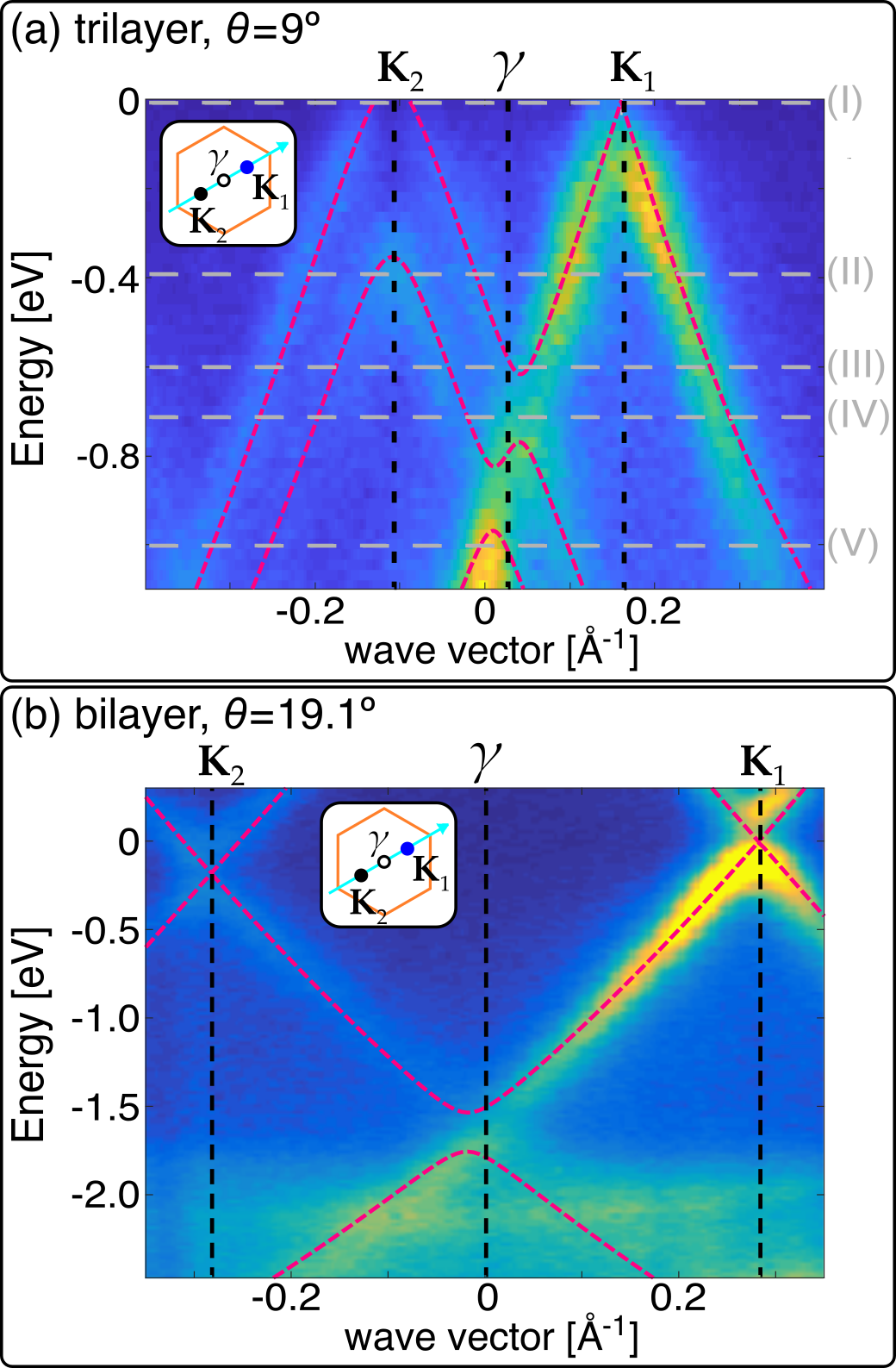}
	\caption{\textbf{Modelling stacks with different twists and layer numbers.} Comparison of the ARPES intensity and the calculated electronic band structure (obtained using the parameter set in Table \ref{table} and shown with red dashed lines) for (a) twisted trilayer with $\theta=9^\circ$ and (b) twisted bilayer with $\theta=19.1^\circ$, both measured along the direction connecting Brillouin zone corners $\vect{K}_{\mathrm{2}}$ and $\vect{K}_{\mathrm{1}}$ as shown in Fig.~\ref{fig:lattice}(c) and indicated in the inset. In (a), the grey dashed lines, labelled (I)-(V), indicate energies for which constant-energy ARPES intensity maps are presented in Fig.~(\ref{fig:maps}).}\label{fig:906}
\end{figure}

The miniband spectrum resulting from our model is shown in red dashed lines in Fig.~\ref{fig:arpes}(a), the functions $t(|\vect{r}|,c_{0})$ and $\tilde{t}(|\vect{q}|,c_{0})$ are plotted in Fig.~\ref{fig:arpes}(b) and the corresponding values of the parameters $\gamma_{0}$, $\gamma_{1}$, $\alpha_{\uppi}$ and $\alpha_{\upsigma}$ are summarized in Table \ref{table}. The interatomic potential we obtain decays more rapidly in the real space (and hence slower in the reciprocal space) than suggested by computational results \cite{laissardiere_nanolett_2010}. Importantly, parametrization of $t(\vect{r},z)$ does not depend on the twist angle and so should be applicable to other graphene stacks with twisted interfaces. It also does not depend on the doping level because, for the relevant range of electric fields, the electrostatic energies $\Delta$ and $u$ do not modify the electron hoppings. At the same time, once these energies are determined for a particular stack, their influence on the band structure (shifting of the positions and magnitudes of anticrossings) is captured through the Hamiltonian $\op{H}$. To confirm applicability of a single parametrization of $t(\vect{r},z)$ to different graphene stacks, we compare in Fig.~\ref{fig:906} the miniband spectra computed using the parameters from Table \ref{table} to ARPES intensities measured along a similar $\vect{K}_{\mathrm{2}}$-$\vect{K}_{\mathrm{1}}$ $k$-space cut for, in Fig.~\ref{fig:906}(a), a trilayer with $\theta=9^{\circ}$ and, in Fig.~\ref{fig:906}(b), twisted bilayer with $\theta=19.1^{\circ}$. Our model describes the bands of both of the structures well, despite changes in the twist angle, number of layers, potentials $u$ and $\Delta$ (which vary with growth conditions and thickness of the stack \cite{peng_advmat_2017} and are determined for each structure individually) and the magnitudes of minigaps. 

\begin{figure*}[t]
    \centering
    \includegraphics[width = 1\linewidth]{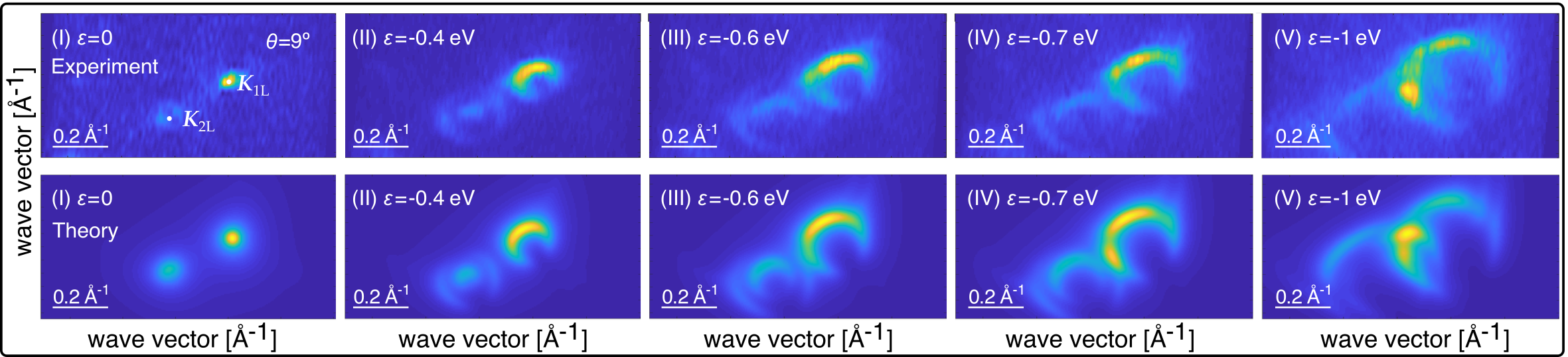}
    \caption{\textbf{Wave function symmetry in ARPES.} Comparison of experimental (top row) and theoretical (bottom row) constant-energy ARPES intensity maps for twisted trilayer graphene with $\theta=9^{\circ}$ for energies indicated with grey dashed lines in Fig.~\ref{fig:906}. The intensities are normalised in each row.}\label{fig:maps}
\end{figure*}

{\bf Probing electron wave function} We assess the accuracy of our parametrization of the interatomic potential, $t(\vect{r},z)$, further by modelling directly the ARPES intensity data (we use approach developed in Ref.~\onlinecite{mucha-kruczynski_prb_2008} and applied to the graphene/hexagonal boron nitride heterostructure in Ref.~\onlinecite{mucha-kruczynski_prb_2016}; see Methods and Supplementary Note 3 for further details). In graphene materials, interference of electrons emitted from different atomic sites within the unit cell provides additional information about the electronic wave function \cite{mucha-kruczynski_prb_2008}. This is best visualized by ARPES intensity patterns at constant electron energy, which we present, both as obtained experimentally (top row) and simulated theoretically (bottom row), in Fig.~\ref{fig:maps} for the trilayer sample with $\theta=9^{\circ}$ and energies indicated with grey dashed lines in Fig.~\ref{fig:906}. For the map at the energy $\varepsilon=0$, the two spots of high intensity indicate the positions of the valleys $\vect{K}_{\mathrm{1}}$ and $\vect{K}_{\mathrm{2}}$. For energies $0<\varepsilon<-0.6$ eV, the bilayer and monolayer dispersions are effectively uncoupled. The crescent-like intensity pattern in the vicinity of $\vect{K}_{\mathrm{1}}$ reflects the pseudospin of $n=1$ (evidence of Berry phase of $\uppi$ \cite{zhang_nature_2005}) of electrons in monolayer graphene. In contrast, in bilayer graphene, the low-energy band hosts massive chiral fermions \cite{novoselov_natphys_2006} with pseudospin $n=2$ so that the outer ring pattern in the vicinity of $\vect{K}_{\mathrm{2}}$ displays two intensity maxima, feature best visible in panel (II). Because in our model all electron hoppings are generated naturally by $t(\vect{r},z)$, agreement of our ARPES simulation with experimental data provides confirmation that our model and parametrization of the interatomic coupling $t(\vect{r},z)$ leads to the correct band structure. Finally, panels (III)-(V) in Fig.~\ref{fig:maps} show the constant-energy maps in the vicinity of the minigaps which open due to hybridization of the bilayer and monolayer bands. The merging of 1L and 2L contours in panel (III) leads to a van Hove singularity and an associated peak in the electronic density of states, similarly to the case of twisted bilayer graphene \cite{ohta_prl_2012} and discussed also for twisted trilayer graphene \cite{peng_advmat_2017} (in the latter, the position of the van Hove singularity is established by tracking the minigap; the former is caused by saddle points in the electronic dispersion as the bands flatten at the anticrossings and so every minigap is accompanied by a van Hove singularity). Overall, our simulated patterns correctly reflect the evolution of the minigap as a function of energy and wave vector as well as the measured photocurrent intensity.

\section{Discussion}

Our parametrization of $t(\vect{r},z)$ is applicable to a wide range of twist angles, including the magic-angle regime \cite{cao_nature_2018, bistritzer_pnas_2011} as well as the $30^{\circ}$-twisted bilayer graphene quasicrystal \cite{ahn_science_2018, yao_pnas_2018}. To mention, it yields the $k$-space interlayer coupling at the graphene Brillouin zone corner $\vect{K}$, $\tilde{t}(|\vect{K}|,c_{0})=0.11$ eV. This agrees with the values used in effective models of the low-twist limit of twisted bilayer graphene \cite{santos_prl_2007, bistritzer_pnas_2011, jung_prb_2014, koshino_njp_2015} which require $\tilde{t}(|\vect{K}|,c_{0})$ as the only parameter. Overall, our form of $t(\vect{r},z)$ decays more rapidly in the real space (and hence slower in the reciprocal space) than usually assumed. This might explain the discrepancy between theory and experimental ARPES intensities of Dirac cone replicas observed for the case of $30^{\circ}$-twisted bilayer graphene in Ref.~\onlinecite{ahn_science_2018}. 

As we have shown, the same interatomic coupling $t(\vect{r},z)$ can be used in graphene structures with different number of layers as, similarly to the case of perfect graphite and other layered materials, coupling to the nearest layer dominates the interlayer couplings. The continuum approach has been applied extensively to model the graphene/graphene interface, including to predict the existence of the magic angle \cite{bistritzer_pnas_2011}. Hence, in Supplementary Figure 1, we use our results to simulate ARPES spectra for twist angles in the vicinity of the magic angle, $\theta\approx 1.1^{\circ}$, and show qualitative agreement with the recent experimental data \cite{utama_arxiv_2019, lisi_arxiv_2020}. The continuum model was also used successfully to interpret experimental observations in graphene on hexagonal boron nitride \cite{ponomarenko_nature_2013} as well as homo- and heterobilayers of transition metal dichalcogenides \cite{alexeev_nature_2019, tran_nature_2019}. Our approach allows for experimental parametrization of the interatomic coupling $t(\vect{r},z)$ for each of these interfaces as well as for others for which influence of neighbouring crystals can be approximated by considering the harmonics of the moir\'{e} potential \cite{wallbank_prb_2013_2, jung_prb_2014, yu_prl_2015, tong_natphys_2017, ruiz-tijerina_prb_2019, wu_prl_2019}. To comment, previous studies suggest that adapting our model to stacks of transition metal dichalcogenides requires taking into account changes in the interlayer distance as a function of the twist angle \cite{yeh_nanolett_2016}. Moreover, in contrast to graphene, for which the part of $\tilde{t}(\vect{q},z)$ most relevant to modelling twisted interfaces is that for $\vect{q}$ pointing to the Brillouin zone corner, $\vect{q}\approx\vect{K}$, for transition metal dichalcogenides more significant changes due to interlayer coupling occur in the vicinity of the $\Gamma$ point. In multilayers of 2H semiconducting dichalcogenides MX$_{2}$ ($\mathrm{M}=\mathrm{Mo,W}$, and $\mathrm{X}=\mathrm{S,Se}$), coupling of the degenerate states at the $\Gamma$ point built of transition metal $d_{z^{2}}$ and chalcogen $p_{z}$ orbitals leads to their hybridization and splitting which drives the direct-to-indirect band gap transition \cite{jin_prl_2013, zhang_naturenano_2014}. Using the form of $t(\vect{r},z)$ suggested in Ref.~\cite{fang_prb_2015} for chalcogen $p_{z}$-to-$p_{z}$ hopping (which dominates the interlayer coupling) in transition metal disulfides and diselenides, we computed the corresponding $\tilde{t}(\vect{q},z)$ and obtained an estimate of $\tilde{t}(\Gamma,c_{\mathrm{X}-\mathrm{X}})\sim 1.2$ eV for interlayer nearest neighbour distance between chalcogen sites, $c_{\mathrm{X}-\mathrm{X}}\approx 3$ \AA. Taking into account the fractional contribution of the $p_{z}$ orbitals to the top valence band states at $\Gamma$ in a monolayer \cite{fang_prb_2015}, we obtain coupling between two such states in bilayer $\sim 0.4$ eV. This, in turn, suggests band splitting of $\sim 0.8$ eV, in qualitative agreement with observations \cite{jin_prl_2013, zhang_naturenano_2014, roldan_annphys_2014}. This supports the idea that our model can accurately describe and parametrize interatomic coupling between materials other than graphene.   

Experimentally, our approach requires fabrication of trilayer (or thicker) stacks with one twisted and one perfect interface in order to benefit from the complementarity of the information obtained from self-consistent real- and momentum-space description of the interfaces. However, to note, building on the observations of superconductivity in magic-angle twisted bilayer graphene \cite{cao_nature_2018, cao_nature_2018_2}, structures containing both a twisted and a perfect interface like twisted trilayer graphene \cite{chen_arxiv_2020, shi_arxiv_2020}, double bilayer graphene \cite{liu_arxiv_2019, burg_prl_2019, shen_arxiv_2019, cao_arxiv_2019, he_arxiv_2020, rickhaus_arxiv_2020} or double bilayer WSe$_{2}$ \cite{an_arxiv_2019} recently attracted attention on its own due to observation of correlated electronic behaviour. Our approach provides one of the avenues to build an experimentally validated single-particle base to study such effects. It could be, in principle, also applied to stacks of different materials, as long as one of the interfaces is commensurate and can be described in the real space in a tight-binding-like fashion. Finally, apart from continuum models, the interatomic coupling $t(\vect{r},z)$ can also be used directly in large scale tight-binding calculations for commensurate twist angles \cite{laissardiere_nanolett_2010, fang_prb_2015, nam_prb_2017, lin_prb_2019, zhao_prl_2020}.

\section{Methods}

{\bf ARPES measurements}
The ARPES measurements were performed at the Spectromicroscopy beamline at the Elettra synchrotron (Trieste, Italy). Before measurements, the samples were annealed at $350^{\circ}$ for 30 minutes. The experiment was then performed at a base pressure of $10^{-10}$ mbar in ultrahigh vacuum and at the temperature of 110 K. We used photons with energy of 74 eV and estimate our energy and angular resolution as 50 meV and $0.5^{\circ}$, respectively. For each sample, we determined the twist angle $\theta$ by measuring the distance between the Brillouin zone corners $\vect{K}_{\mathrm{2}}$ and $\vect{K}_{\mathrm{1}}$ which depends on the twist angle, $|\vect{K}_{\mathrm{2}}-\vect{K}_{\mathrm{1}}|=\tfrac{8\uppi}{3\sqrt{3}|\vect{r}_{AB}|}\sin\tfrac{\theta}{2}$. Further comments on experimental analysis of ARPES intensity are provided in Supplementary Note 4.

{\bf Theoretical calculations} We write the Hamiltonian $\op{H}$ in Equation \eqref{ttlg} in the basis of sublattice Bloch states constructed of carbon $p_{z}$ orbitals $\phi(\vect{r}_{\mathrm{3D}})$ \cite{castro_neto_rmp_2009}, 
\begin{align*}
\Ket{\vect{k},X}_l = \frac{1}{\sqrt{N}}\sum_{\vect{R}_{l}} \mathrm{e}^{\mathrm{i}\vect{k}\cdot(\vect{R}_{l}+\vect{\tau}_{X,l})} \phi(\vect{r}_{\mathrm{3D}}-\vect{R}_{l}-\vect{\tau}_{X,l}),
\end{align*}
where $\vect{k}$ is electron wave vector, $X=A,B$ is the sublattice, $\vect{R}_{l}$ are the lattice vectors of layer $l$ and $\vect{\tau}_{X,l}$ points to the site $X$ in layer $l$ within the unit cell selected by $\vect{R}_{l}$. We include in the basis all states coupled to $\vect{k}$ through $\op{T}(\theta)$ which are less than a distance $\tfrac{28\uppi}{3\sqrt{3}r_{AB}}\sin\tfrac{\theta}{2}$ away from it, compute the matrix elements of $\op{H}$ in this truncated basis and diagonalize the resulting matrix numerically. In order to simulate the ARPES intensity, we project the eigenstates of the moir\'{e} Hamiltonian, $\op{H}$, on a plane-wave-like final state (see Supplementary Note 3 for more details and Ref.~\onlinecite{mucha-kruczynski_prb_2016} for a detailed discussion of this approach for the case of graphene on hexagonal boron nitride). We determine the broadening of the ARPES signal as well as the decay constant for the intensity of Bernal bilayer signal by fitting to the experimental data.

\section{Acknowledgements}

J.J.P.T.~was supported by EPSRC through the University of Bath Doctoral Training Partnership, EPSRC Grant No.~EP/M507982/1. N.C.~was supported by the Institute for Mathematical Innovation at the University of Bath. M.M.-K.~acknowledges funding from the University of Bath International Research Funding Scheme. D.P.~and H.P.~acknowledge support from the China Scholarship Council.

\section{Author contributions}

H.P.~and N.S.~carried out the ARPES measurements with the assistance of A.B.~and Y.C. D.P.~analyzed the ARPES data. H.W.~and H.L.P.~grew the samples. J.J.P.T.~and M.M.-K.~built the theoretical model. J.J.P.T.~and N.C.~performed the miniband and ARPES simulations. M.M.-K.~conceived the project and supervised the theoretical analysis. J.J.P.T.~and M.M.-K.~wrote the manuscript with input from D.P., A.B.~and Y.C.

\section{Competing Interests} The authors declare no competing interests.

\section{Data availability}
The data used in this study are available from the University of Bath data archive at \href{https://doi.org/10.15125/BATH-00864}{https://doi.org/10.15125/BATH-00864} \cite{dataset}.


\begin{thebibliography}{99}

\bibitem{novoselov_science_2004} Novoselov,  K.~S.~et al. Electric field effect in atomically thin carbon films. \href{https://dx.doi.org/10.1126/science.1102896}{{\em Science} {\bf 306}, 666-669 (2004)}.

\bibitem{geim_nature_2013} Geim, A.~K.~\& Grigorieva, I.~V. Van der Waals heterostructures. \href{https://dx.doi.org/10.1038/nature12385}{{\em Nature} {\bf 499}, 419-425 (2013)}.

\bibitem{novoselov_science_2016} Novoselov, K.~S., Mishchenko, A., Carvalho, A.~\& Castro Neto, A.~H. 2D materials and van der Waals heterostructures. \href{https://dx.doi.org/10.1126/science.aac9439}{{\em Science} {\bf 353}, aac9439 (2016)}.

\bibitem{liu_natrevmat_2016} Liu, Y., Weiss, N.~O., Duan, X., Cheng, H.-C., Huang, Y.~\& Duan, X. Van der Waals heterostructures and devices. \href{https://dx.doi.org/10.1038/natrevmats.2016.42}{{\em Nature Reviews Materials} {\bf 1}, 16042 (2016)}.

\bibitem{ponomarenko_nature_2013} Ponomarenko, L.~A.~et al. Cloning of dirac fermions in graphene superlattices. \href{https://dx.doi.org/10.1038/nature12187}{{\em Nature} {\bf 497}, 594-597 (2013)}.

\bibitem{dean_nature_2013} Dean, C.~R.~et al. Hofstadter's butterfly and the fractal quantum hall effect in moir{\'e} superlattices. \href{https://dx.doi.org/10.1038/nature12186}{{\em Nature} {\bf 497}, 598-602 (2013)}.

\bibitem{chen_nanolett_2018} Chen, C.~et al. Emergence of interfacial polarons from electron-phonon coupling in graphene/h-BN van der Waals heterostructures. \href{https://dx.doi.org/10.1021/acs.nanolett.7b04604}{{\em Nano Letters} {\bf 18}, 1082-1087 (2018)}.

\bibitem{fang_pnas_2014} Fang, H.~et al. Strong interlayer coupling in van der waals heterostructures built from single-layer chalcogenides. \href{https://dx.doi.org/10.1073/pnas.1405435111}{{\em Proceedings of the National Academy of Sciences} {\bf 111}, 6198-6202 (2014)}.

\bibitem{rivera_natnano_2018} Rivera, P.~et al. Interlayer valley excitons in heterobilayers of transition metal dichalcogenides. \href{https://dx.doi.org/10.1038/s41565-018-0193-0}{{\em Nature Nanotechnology} {\bf 13}, 1004-1015 (2018)}.

\bibitem{cao_nature_2018} Cao, Y.~et al. Correlated insulator behaviour at half-filling in magic-angle graphene superlattices. \href{https://dx.doi.org/10.1038/nature26154}{{\em Nature} {\bf 556}, 80-84 (2018)}.

\bibitem{cao_nature_2018_2} Cao, Y.~et al. Unconventional superconductivity in magic-angle graphene superlattices. \href{https://dx.doi.org/10.1038/nature26160}{{\em Nature} {\bf 556}, 43-50 (2018)}.

\bibitem{chari_nanolett_2016} Chari, T., Ribeiro-Palau, R., Dean, C.~R. \& Shepard, K. Resistivity of Rotated Graphite-Graphene Contacts. \href{https://dx.doi.org/10.1021/acs.nanolett.6b01657}{{\em Nano Letters} {\bf  16}, 4477-4482 (2016)}.

\bibitem{ribeiro-palau_science_2018} Ribeiro-Palau, R.~et al. Twistable electronics with dynamically rotatable heterostructures. \href{https://dx.doi.org/10.1126/science.aat6981}{{\em Science} {\bf 361}, 690-693 (2018)}.

\bibitem{finney_natnano_2019} Finney, N.~R.~et al. Tunable crystal symmetry in graphene-boron nitride heterostructures with coexisting moir\'{e} superlattices. \href{https://dx.doi.org/10.1038/s41565-019-0547-2}{{\em Nature Nanotechnology} {\bf 14}, 1029-1034 (2019)}.

\bibitem{ohta_prl_2012} Ohta, T.~et al. Evidence for Interlayer Coupling and Moir\'{e} Periodic Potentials in Twisted Bilayer Graphene. \href{https://dx.doi.org/10.1103/PhysRevLett.109.186807}{{\em Physical Review Letters} {\bf 109}, 186807 (2012)}.
% ARPES of twisted bilayer graphene van Hove singularity

\bibitem{diaz_nanolett_2015} Diaz, H.~C.~et al. Direct Observation of Interlayer Hybridization and Dirac Relativistic Carriers in Graphene/MoS$_{2}$ van der Waals Heterostructures. \href{https://dx.doi.org/10.1021/nl504167y}{{\em Nano Letters} {\bf 15}, 1135-1140 (2015)}.

\bibitem{wilson_sciadv_2017} Wilson, N.~R.~et al. Determination of band offsets, hybridization, and exciton binding in 2D semiconductor heterostructures. \href{https://dx.doi.org/10.1126/sciadv.1601832}{{\em Science Advances} {\bf 3}, e1601832 (2017)}.

\bibitem{pierucci_nanolett_2016} Pierucci, D.~et al. Band Alignment and Minigaps in Monolayer MoS$_{2}$-Graphene van der Waals Heterostructures. \href{https://dx.doi.org/10.1021/acs.nanolett.6b00609}{{\em Nano Letters} {\bf 16}, 4054-4061 (2016)}.

\bibitem{ulstrup_sciadv_2020} Ulstrup, S. et al. Direct observation of minibands in a twisted graphene/WS$_{2}$ bilayer. \href{https://dx.doi.org/10.1126/sciadv.aay6104}{{\em Science Advances} {\bf 6}, eaay6104 (2020)}.

\bibitem{yeh_nanolett_2016} Yeh, P.-C.~et al. Direct Measurement of the Tunable Electronic Structure of Bilayer MoS$_{2}$ by Interlayer Twist. \href{https://dx.doi.org/10.1021/acs.nanolett.5b03883}{{\em Nano Letters} {\bf 16}, 953-959 (2016)}.
% added post review

\bibitem{zribi_npj2D_2019} Zribi, J.~et al. Strong interlayer hybridization in the aligned SnS$_{2}$/WSe$_{2}$ hetero-bilayer structure. \href{https://dx.doi.org/10.1038/s41699-019-0109-3}{{\em npj 2D Materials and Applications} {\bf 3}, 27 (2019)}.

\bibitem{zande_nanolett_2014} van der Zande, A.~M.~et al. Tailoring the Electronic Structure in Bilayer Molybdenum Disulfide via Interlayer Twist. \href{https://dx.doi.org/10.1021/nl501077m}{{\em Nano Letters} {\bf  14}, 3869-3875 (2014)}.

\bibitem{huang_nanolett_2014} Huang, S.~et al. Probing the Interlayer Coupling of Twisted Bilayer MoS$_{2}$ Using Photoluminescence Spectroscopy. \href{https://dx.doi.org/10.1021/nl5014597}{{\em Nano Letters} {\bf  14}, 5500-5508 (2014)}.

\bibitem{liu_naturecomms_2014} Liu, K.~et al. Evolution of interlayer coupling in twisted molybdenum disulfide bilayers. \href{https://dx.doi.org/10.1038/ncomms5966}{{\em Nature Communications} {\bf 5}, 4966 (2014)}.
% added post review


\bibitem{laissardiere_nanolett_2010} de Laissardiere, G.~T., Mayou, D.~\& Magaud, L. Localization of Dirac Electrons in Rotated Graphene Bilayers. \href{https://dx.doi.org/10.1021/nl902948m}{{\em Nano Letters} {\bf 10}, 804-808 (2010)}.

\bibitem{fang_prb_2015} Fang, S.~et al. Ab initio tight-binding Hamiltonian for transition metal dichalcogenides. \href{https://dx.doi.org/10.1103/PhysRevB.92.205108}{{\em Physical Review B} {\bf 92}, 205108 (2015)}.

\bibitem{santos_prl_2007} Lopes dos Santos, J.~M.~B., Peres, N.~M.~R.~\& Castro Neto, A.~H. Graphene Bilayer with a Twist: Electronic Structure. \href{https://dx.doi.org/10.1103/PhysRevLett.99.256802}{{\em Physical Review Letters} {\bf 99}, 256802 (2007)}.

\bibitem{wallbank_prb_2013} Wallbank, J.~R., Patel, A.~A., Mucha-Kruczynski, M., Geim, A.~K.~\& Fal'ko, V.~I. Generic miniband structure of graphene on a hexagonal substrate. \href{https://dx.doi.org/10.1103/PhysRevB.87.245408}{{\em Physical Review B} {\bf 87}, 245408 (2013)}.

	%%%


	%%%
	
	%%% growth and characterization - methods

\bibitem{peng_advmat_2017} Peng, H.~et al. Substrate Doping Effect and Unusually Large Angle van Hove Singularity Evolution in Twisted Bi- and Multilayer Graphene. \href{https://dx.doi.org/10.1002/adma.201606741}{{\em Advanced Materials} {\bf 29}, 1606741 (2017)}.

\bibitem{mattevi_jmc_2011} Mattevi, C., Kim, H.~\& Chhowalla, M. A review of chemical vapour deposition of graphene on copper. \href{https://dx.doi.org/10.1039/C0JM02126A}{{\em Journal of Materials Chemistry} {\bf 21}, 3324-3334 (2011)}.

	%%%

\bibitem{castro_neto_rmp_2009} Castro Neto, A.~H., Guinea, F., Peres, N.~M.~R., Novoselov, K.~S.~\& Geim, A.~K. The electronic properties of graphene. \href{https://dx.doi.org/10.1103/RevModPhys.81.109}{{\em Rev. Mod. Phys.} {\bf 81}, 109-162 (2009)}.

\bibitem{koshino_prb_2009} Koshino, M.~\& McCann, E. Gate-induced interlayer asymmetry in ABA-stacked trilayer graphene. \href{https://dx.doi.org/10.1103/PhysRevB.79.125443}{{\em Phys. Rev. B} {\bf 79}, 125443 (2009)}.

\bibitem{mccann_prl_2006} McCann, E.~\& Fal'ko, V.~I. Landau-level degeneracy and quantum hall effect in a graphite bilayer. \href{https://dx.doi.org/10.1103/PhysRevLett.96.086805}{{\em Phys. Rev. Lett.} {\bf 96}, 086805 (2006)}.

\bibitem{bistritzer_pnas_2011} Bistritzer, R.~\& MacDonald, A.~H. Moir{\'e} bands in twisted double-layer graphene. \href{https://dx.doi.org/10.1073/pnas.1108174108}{{\em Proceedings of the National Academy of Sciences} {\bf 108}, 12233-12237 (2011)}.

\bibitem{koshino_njp_2015} Koshino, M. Interlayer interaction in general incommensurate atomic layers. \href{https://dx.doi.org/10.1088/1367-2630/17/1/015014}{{\em New Journal of Physics} {\bf 17}, 015014 (2015)}.

\bibitem{koster_physrev_1954} Slater, J.~C.~\& Koster, G.~F. Simplified LCAO Method for the Periodic Potential Problem. \href{https://dx.doi.org/10.1103/PhysRev.94.1498}{{\em Physical Review} {\bf 94}, 1498-1524 (1954)}.

\bibitem{mucha-kruczynski_prb_2008} Mucha-Kruczynski, M.~et al. Characterization of graphene through anisotropy of constant-energy maps in angle-resolved photoemission. \href{https://dx.doi.org/10.1103/PhysRevB.77.195403}{{\em Physical Review B} {\bf 77}, 195403 (2008)}.

\bibitem{mucha-kruczynski_prb_2016} Mucha-Kruczynski, M., Wallbank, J.~R.~\& Fal'ko, V.~I. Moir\'e miniband features in the angle-resolved photoemission spectra of graphene/$h\mathrm{BN}$ heterostructures. \href{https://dx.doi.org/10.1103/PhysRevB.93.085409}{{\em Physical Review B} {\bf 93}, 085409 (2016)}.

\bibitem{zhang_nature_2005} Zhang, Y., Tan, Y.-W., Stormer, H.~L.~\& Kim, P. Experimental observation of the quantum Hall effect and Berry's phase in graphene. \href{https://dx.doi.org/10.1038/nature04235}{{\em Nature} {\bf 438}, 201-204 (2005)}.

\bibitem{novoselov_natphys_2006} Novoselov, K.~S.~et al. Unconventional quantum Hall effect and Berry's phase of $2\uppi$ in bilayer graphene. \href{https://dx.doi.org/10.1038/nphys245}{{\em Nature Physics} {\bf 2}, 177-180 (2006)}. % bilayer

	% Quasi-crystal ARPES
\bibitem{ahn_science_2018} Ahn, S.~J.~et al. Dirac electrons in a dodecagonal graphene quasicrystal. \href{https://dx.doi.org/10.1126/science.aar8412}{{\em Science} {\bf 361}, 782-786 (2018)}.

\bibitem{yao_pnas_2018} Yao, W.~et al. Quasicrystalline $30^{\circ}$ twisted bilayer graphene as an incommensurate superlattice with strong interlayer coupling. \href{https://dx.doi.org/10.1073/pnas.1720865115}{{\em Proceedings of the National Academy of Sciences} {\bf 115}, 6928-6933 (2018)}. 

	%%%
	
\bibitem{jung_prb_2014} Jung, J., Raoux, A., Qiao, Z.~\& MacDonald, A.~H. Ab initio theory of moir\'{e} superlattice bands in layered two-dimensional materials. \href{https://dx.doi.org/10.1103/PhysRevB.89.205414}{{\em Physical Review B} {\bf 89}, 205414 (2014)}.

\bibitem{utama_arxiv_2019} Utama, M.~I.~B.~et al. Visualization of the flat electronic band in twisted bilayer graphene near the magic angle twist. Preprint at \href{https://arxiv.org/abs/1912.00587}{arXiv:1912.00587} (2019).
% experiment - ARPES of small-angle twisted blayer

\bibitem{lisi_arxiv_2020} Lisi, S.~et al. Direct evidence for flat bands in twisted bilayer graphene from nano-ARPES. Preprint at \href{https://arxiv.org/abs/2002.02289}{arXiv:2002.02289} (2020). 
% experiment - ARPES of small-angle twisted bilayer

\bibitem{alexeev_nature_2019} Alexeev, E.~M.~et al. Resonantly hybridized excitons in moir\'{e} superlattices in van der Waals heterostructures. \href{https://dx.doi.org/10.1038/s41586-019-0986-9}{{\em Nature} {\bf 567}, 81-86 (2019)}.

\bibitem{tran_nature_2019} Tran, K.~et al. Evidence for moir\'{e} excitons in van der Waals heterostructures. \href{https://dx.doi.org/10.1038/s41586-019-0975-z}{{\em Nature} {\bf 567}, 71-75 (2019)}.

\bibitem{wallbank_prb_2013_2} Wallbank, J.~R., Mucha-Kruczynski, M.~\& Fal'ko, V.~I. Moir\'{e} minibands in graphene heterostructures with almost commensurate $\sqrt{3}\times\sqrt{3}$ hexagonal crystals. \href{https://dx.doi.org/10.1103/PhysRevB.88.155415}{{\em Physical Review B} {\bf 88}, 155415 (2013)}.

\bibitem{yu_prl_2015} Yu, H., Wang, Y., Tong, Q., Xu, X.~\& Yao, W. Anomalous Light Cones and Valley Optical Selection Rules of InterlayerExcitons in Twisted Heterobilayers. \href{https://dx.doi.org/10.1103/PhysRevLett.115.187002}{{\em Physical Review Letters} {\bf 115}, 187002 (2015)}.

\bibitem{tong_natphys_2017} Tong, Q.~et al. Topological mosaics in moir\'{e} superlattices of van der Waals heterobilayers. \href{https://dx.doi.org/10.1038/nphys3968}{{\em Nature Physics} {\bf 13}, 356-362 (2017)}.

\bibitem{ruiz-tijerina_prb_2019} Ruiz-Tijerina, D.~A.~\& Fal'ko, V.~I. Interlayer hybridization and moir\'{e} superlattice minibands for electrons and excitons in heterobilayers of transition-metal dichalcogenides. \href{https://dx.doi.org/10.1103/PhysRevB.99.125424}{{\em Physical Review B} {\bf 99}, 125424 (2019)}.

\bibitem{wu_prl_2019} Wu, F., Lovorn, T., Tutuc, E., Martin, I.~\& MacDonald, A.~H. Topological Insulators in Twisted Transition Metal Dichalcogenide Homobilayers. \href{https://dx.doi.org/10.1103/PhysRevLett.122.086402}{{\em Physical Review Letters} {\bf 122}, 086402 (2019)}.
	%%%

\bibitem{jin_prl_2013} Jin, W.~et al. Direct Measurement of the Thickness-Dependent Electronic Band Structure of MoS$_{2}$ Using Angle-Resolved Photoemission Spectroscopy. \href{https://dx.doi.org/10.1103/PhysRevLett.111.106801}{{\em Physical Review Letters} {\bf 111}, 106801 (2013)}.

\bibitem{zhang_naturenano_2014} Zhang, Y.~et al. Direct observation of the transition from indirect to direct bandgap in atomically thin epitaxial MoSe$_{2}$. \href{https://dx.doi.org/10.1038/nnano.2013.277}{{\em Nature Nanotechnology} {\bf 9}, 111-115 (2014)}.

\bibitem{roldan_annphys_2014} Roldan, R.~et al. Electronic properties of single-layer and multilayer transition metal dichalcogenides MX$_{2}$ ($\mathrm{M}=\mathrm{Mo, W}$ and $\mathrm{X}=\mathrm{S, Se}$). \href{https://dx.doi.org/10.1002/andp.201400128}{{\em Annalen der Physik} {\bf 526}, 347-357 (2014)}.


	% twisted trilayer graphene

\bibitem{chen_arxiv_2020} Chen, S.~et al. Electrically tunable correlated and topological states in twisted monolayer-bilayer graphene. Preprint at \href{https://arxiv.org/abs/2004.11340}{arXiv:2004.11340} (2020).
% experiment - (1+2)

\bibitem{shi_arxiv_2020} Shi, Y.~et al. Tunable van Hove Singularities and Correlated States in Twisted Trilayer Graphene. Preprint at \href{https://arxiv.org/abs/2004.12414}{arXiv:2004.12414} (2020). 
% experiment - (1+2)

	%%%

	% double bilayer graphene
	
\bibitem{liu_arxiv_2019} Liu, X.~et al. Spin-polarized Correlated Insulator and Superconductor in Twisted Double Bilayer Graphene. Preprint at \href{https://arxiv.org/abs/1903.08130}{arXiv:1903.08130} (2019).
% experiment - double bilayer

\bibitem{burg_prl_2019} Burg, G.~W.~et al. Correlated Insulating States in Twisted Double Bilayer Graphene. \href{https://dx.doi.org/10.1103/PhysRevLett.123.197702}{{\em Physical Review Letters} {\bf 123}, 197702 (2019)}.
% experiment - double-bilayer

\bibitem{shen_arxiv_2019} Shen, C.~et al. Correlated states in twisted double bilayer graphene. \href{https://doi.org/10.1038/s41567-020-0825-9}{{\em Nature Physics} {\bf 16}, 520-525 (2020)}. % double bilayer graphene

\bibitem{cao_arxiv_2019} Cao, Y.~et al. Tunable correlated states and spin-polarized phases in twisted bilayer-bilayer graphene. \href{https://doi.org/10.1038/s41586-020-2260-6}{{\em Nature} {\bf 583}, 215-220 (2020)}.
% experiment - double-bilayer

\bibitem{he_arxiv_2020} He, M.~et al. Tunable correlation-driven symmetry breaking in twisted double bilayer graphene. Preprint at \href{https://arxiv.org/abs/2002.08904}{arXiv:2002.08904} (2020). % double bilayer graphene

\bibitem{rickhaus_arxiv_2020} Rickhaus, P.~et al. Density-Wave States in Twisted Double-Bilayer Graphene. Preprint at \href{https://arxiv.org/abs/2005.05373}{arXiv:2005.05373} (2020).

	%%%

	% WSe2 heterostructures

\bibitem{an_arxiv_2019} An, L.~et al. Interaction effects and superconductivity signatures in twisted double-bilayer WSe2. Preprint at \href{https://arxiv.org/abs/1907.03966}{arXiv:1907.03966} (2019). % double bilayer WSe2

	%%%

\bibitem{nam_prb_2017} Nam, N.~N.~T.~\& Koshino, M. Lattice relaxation and energy band modulation in twisted bilayer graphene. \href{https://doi.org/10.1103/PhysRevB.96.075311}{{\em Physical Review B} {\bf 96}, 075311 (2017)}.

\bibitem{lin_prb_2019} Lin, X.~\& Tomanek, D. Minimum model for the electronic structure of twisted bilayer graphene and related structures. \href{https://doi.org/10.1103/PhysRevB.98.081410}{{\em Physical Review B} {\bf 98}, 081410 (2019)}.

\bibitem{zhao_prl_2020} Zhao, X.-J., Yang, Y., Zhang, D.-B.~\& Wei, S.-H. Formation of Bloch Flat Bands in Polar Twisted Bilayers without Magic Angles. \href{https://doi.org/10.1103/PhysRevLett.124.086401}{{\em Physical Review Letters} {\bf 124}, 086401 (2020)}.

\bibitem{dataset} Thompson, J.~J.~P. Dataset for article "Determination of interatomic coupling between two-dimensional crystals using angle-resolved photoemission spectroscopy" by Thompson et al. Bath: University of Bath Research Data Archive. \href{https://doi.org/10.15125/BATH-00864}{https://doi.org/10.15125/BATH-00864}

\end{thebibliography}
\end{document}

%\beginsupplement
	
\title{Supplementary Information: \\ Determination of interatomic coupling in twisted graphene using angle-resolved photoemission spectroscopy}

\author{Thompson et al.}

\maketitle

%\tableofcontents

\newpage

%\section{Electronic minibands of twisted trilayer graphene}
%\begin{center}
\begin{flushleft}
{\bf Supplementary Note 1: Electronic minibands of twisted trilayer graphene}
%\end{center}
\end{flushleft}

%	\subsection{Hamiltonian}

As mentioned in the main text, we aim to construct our Hamiltonian $\op{H}$ based on blocks corresponding to intra- and interlayer couplings,
\begin{align} \label{ttlg}
\op{H}= \begin{pmatrix}
      \op{H}_0\left(0, \Delta-u \right) & \op{T}(0) & 0 \\
      \op{T}^\dagger(0)& \op{H}_0\left(0,\Delta +u \right)&\op{T}(\theta) \\
      0& \op{T}^\dagger(\theta) & \op{H}_0\left(\theta,0\right)
      \end{pmatrix}.
\end{align}
Above, the diagonal blocks $\op{H}_0\left(\theta_i, \varepsilon_i\right)$ describe consecutive graphene layers (starting from the top left for layer 1) at a twist angle $\theta_i$ and with on-site energies of atomic sites in this layer, $\varepsilon_i$. Without loss of generality, we choose $\theta_{1}=0$ (we measure twists with respect to the crystallographic directions of layer 1) and, from our structure, $\theta_{2}=0$ immediately follows. In contrast, we set the reference point for energy as the on-site energies of layer 3 equal to zero, $\varepsilon_3=0$. From the definitions of the energies $\Delta$ and $u$ in the main text, we get
\begin{align*}
\varepsilon_1=\Delta-u,\,\,\,\,\,\,\varepsilon_2=\Delta+u.
\end{align*} 

Consider now the sublattice Bloch states constructed of carbon $p_{z}$ orbitals $\phi(\vect{r}_{\mathrm{3D}})\equiv\phi(\vect{r},z)$, 
\begin{align}\label{eqn:Bloch}
\Ket{\vect{k},X}_l = \frac{1}{\sqrt{N}}\sum_{\vect{R}_{l}} e^{i\vect{k}\cdot(\vect{R}_{l}+\vect{\tau}_{X,l})} \phi(\vect{r}-\vect{R}_{l}-\vect{\tau}_{X,l},z-z_{l}),
\end{align}
where $\vect{k}=(k_{x},k_{y})$ is electron wave vector, $X=A,B$ is the sublattice, $\vect{R}_{l}$ are the lattice vectors of layer $l$, $\vect{\tau}_{X,l}$ points to the site $X$ in layer $l$ within the unit cell selected by $\vect{R}_{l}$ and $z_{l}$ defines the position of layer $l$ along the $z$-axis. In the basis $\{\Ket{\vect{k},A}_l,\Ket{\vect{k},B}_l\}$, the intralayer blocks take a familiar tight-binding form  \cite{castro_neto_rmp_2009},
\begin{align}
      \op{H}_{0}(\theta_{l}, u_{l}) &=  \begin{pmatrix}
    u_{l}& -\gamma_0 f(\op{R}_{\theta_{l}}\vect{k})\\ 
       -\gamma_0 f^*(\op{R}_{\theta_{l}}\vect{k}) & u_{l}
      \end{pmatrix}, 
      \\
      f(\vect{k}) &= \exp \left( \dfrac{i k_y a}{\sqrt{3}}\right) + 2 \exp \left(-\dfrac{i k_y a}{2 \sqrt{3}}\right) \cos\left(\dfrac{k_x a}{2}\right),\nonumber
\end{align}
where $a=\sqrt{3}|\vect{r}_{AB}|$ is the lattice constant of graphene. Similarly, the interlayer coupling block between layers 1 and 2 which are perfectly aligned and stacked according to the Bernal stacking \cite{mccann_prl_2006},
\begin{align}\label{eqn:Bernal}
\op{T}(0)= \begin{pmatrix} 0 & 0 \\ \gamma_{1} & 0 \end{pmatrix}. 
\end{align}

Let us now consider the interface between layers 2 and 3. The positions of the $j$-th and $m$-th sublattices in these layers can be expressed as
\begin{align}\begin{split}
    &\vect{R}_j = n_1\vect{a}_{1}+n_2\vect{a}_{2}+ \vect{\tau}_{j,2}, \\
    &\vect{R}^{'}_m= {n}^{'}_{1}\op{R}_{\theta}\vect{a}_{1}+{n}^{'}_2\op{R}_{\theta}\vect{a}_{2} +\op{R}_{\theta}\vect{\tau}_{m,3},
\end{split}\end{align}
where $\vect{a}_{1}$ and $\vect{a}_{2}$ are the primitive lattice vectors of layer 2 and $n_1,n_2,{n}^{'}_{1},{n}^{'}_{2}\in\mathbb{Z}$. For no twist, the two layers would be Bernal-stacked and that prescribes the relation between $\vect{\tau}_{j,2}$ and $\vect{\tau}_{m,3}$. Here, we choose
\begin{align}
\begin{matrix*}[l]
\vect{\tau}_{A,2} = -(0,a/\sqrt{3}), & \vect{\tau}_{B,2} = (0,0), \\
\vect{\tau}_{A,3} = (0,0), & \vect{\tau}_{B,3} = (0,a/\sqrt{3}).
\end{matrix*}
\end{align}
Following earlier work \cite{bistritzer_pnas_2011, koshino_njp_2015}, the interlayer coupling matrix element $_{3}\!\braket{\vect{k}',m,|\op{T}|\vect{k},j}_{2}$ between sublattice Bloch states on layers 2 and 3 can be written as
\begin{align}
_{3}\!\braket{\vect{k}',m,|\op{T}|\vect{k},j}_{2} = \sum_{\vect{G},\vect{G}'}\tilde{t}(\vect{k}+\vect{G},c_{0})e^{-i\vect{G}\cdot\vect{\tau}_{j,2}}e^{i\vect{G}'\cdot \op{R}_\theta \vect{\tau}_{m,3}}\,\delta_{\vect{k}+\vect{G},\vect{k}'+\vect{G}'}.
\end{align}
The matrix element above can be understood in the following way: (1) the Kronecker delta term, $\delta_{\vect{k}+\vect{G},\vect{k}'+\vect{G}'}$, expresses conservation of crystal momentum and determines the set of momenta on the top and bottom layers which are coupled (these are effectively momenta $\vect{k}$ and $\vect{k}'$ which are offset by a moir\'{e} reciprocal lattice vector $\vect{g}=\vect{G}'-\vect{G}$); (2) the phase  $ e^{i (\vect{G}'\cdot \op{R}_\theta \vect{\tau}_{m,3}-\vect{G} \cdot \vect{\tau}_{l,2})}$ describes the phase factor associated with the coupling of orbitals $m$ and $l$ as a result of translations by reciprocal lattice vectors in each layer. Qualitatively, these phases describe a continuous transition between regions of $AA$, $AB$ and $BA$-type stacking present in the moir\'{e} pattern between twisted graphene layers. Finally, the strength of the coupling, $\tilde{t}(\vect{k}+\vect{G}, c_0)$, is prescribed by the (total) momentum of the electron tunnelling between the layers 2 and 3.

By writing all four matrix elements in the form of a $2\times 2$ matrix, we obtain Eq.~(2) of the main text,
\begin{align}\label{eqn:twist}
 \op{T}(\theta) = &\sum_{\vect{G}, \vect{G'}}\tilde{t}(\vect{k}+\vect{G},c_0) \times \begin{pmatrix}
 e^{i \vect{G} \cdot \vect{\tau}} &e^{i (\vect{G} +\op{R}_{\theta}\vect{G'}) \cdot \vect{\tau} } \\
 1 & e^{i \op{R}_{\theta}\vect{G'} \cdot \vect{\tau}}
 \end{pmatrix} \delta_{\vect{k}+\vect{G},\vect{k'}+\vect{G'}},
\end{align}
where we only use $\vect{\tau}\equiv\vect{\tau}_{A,2} = -(0,a/\sqrt{3})$.

In order to obtain the energy dispersion for a given $\vect{k}$, we include into our Hamiltonian all states coupled to $\vect{k}$ through $\op{T}(\theta)$ which are less than a distance $\tfrac{28\pi}{3\sqrt{3}r_{AB}}\sin\tfrac{\theta}{2}$ away from it, compute the matrix elements of $\op{H}$ in this truncated basis and diagonalize the resulting matrix numerically.

\newpage
%	\subsection{Self-consistency check for interlayer coupling blocks}
\begin{flushleft}
{\bf Supplementary Note 2: Self-consistency check for interlayer coupling blocks}
\end{flushleft}

We use comparison between the limit $\theta\rightarrow 0$ of the interlayer block $ \op{T}(\theta)$ written in the reciprocal space, Supplementary Equation~\eqref{eqn:twist}, and the $\op{T}(0)$ block corresponding to interlayer coupling for a perfect Bernal bilayer, Supplemetary Equation \eqref{eqn:Bernal}, as a check of self-consistency of our model. In this limit, the superlattice Brillouin zone shrinks to a single point and the reciprocal vectors of both layers become identical.

Now, consider an electron state with wave vector $\vect{k}$ in the vicinity of the valley corner $\vect{K}$, $\vect{k}\approx\vect{K}$, as well as the states with wave vectors $\vect{k}+\vect{G}$. These can be grouped according to their distance from the origin [for example, the first group is given by $\vect{K}$, $\vect{K}-\tfrac{2\pi}{a}(1,\tfrac{1}{\sqrt{3}})$ and $\vect{K}-\tfrac{2\pi}{a}(1,-\tfrac{1}{\sqrt{3}})$]. Hence, the terms in the sum in Supplementary Equation \eqref{eqn:twist} can be grouped according to the same coupling magnitudes, $\tilde{t}(\vect{k}+\vect{G},c_0)\equiv\tilde{t}(|\vect{k}+\vect{G}|,c_0)\approx\tilde{t}(|\vect{K}+\vect{G}|,c_0)$. Because of the rapid decay of $\tilde{t}(\vect{k},c_0)$ as a function of $|\vect{k}|$, the sum converges. Taking note of the partial cancellation of the phase factors, in the limit $\theta\rightarrow 0$ equivalence of the descriptions in the real and reciprocal spaces requires that
\begin{align}
\op{T}(\theta\rightarrow 0) \approx 3\left[ \tilde{t}(\vect{K},c_0)+ \tilde{t}(2\vect{K},c_0) +2\tilde{t}(\sqrt{7}\vect{K},c_0)+...\right]\begin{pmatrix}0 & 0 \\1 & 0\end{pmatrix} = \begin{pmatrix}
0 & 0 \\ \gamma_1 & 0\end{pmatrix},
\end{align}
to reproduce the interlayer coupling in Bernal bilayers. As $\gamma_{1}\equiv t(0,c_{0})$, this condition provides us with additional constraints on the behaviour of $t(\vect{r},z)$ and its Fourier transform $t(\vect{k},z)$.

\newpage
%\section{Theoretical model of ARPES intensity}
\begin{flushleft}
{\bf Supplementary Note 3: Theoretical model of ARPES intensity}
\end{flushleft}

Using Fermi's golden rule, we write ARPES intensity as \cite{mucha-kruczynski_prb_2008, mucha-kruczynski_prb_2016}
\begin{align}
I \propto  \sum_{i}\left| M_{f,i} \right|^2 \delta (\omega+\varepsilon_{i,\vect{k}}-W-\varepsilon_{\vect{p}_{e}}),
\end{align}
where $M_{f,i}$ is the matrix element describing transition of the electron from the initial state in the crystal in band $i$ to the final state $f$, $\omega$ is the energy of the incident photon, $\varepsilon_{i,\vect{k}}$ is the energy of an electron in the crystal in band $i$ and with wave vector $\vect{k}$, $\varepsilon_{\vect{p}_{e}}$ is the energy of the photoelectron with momentum $\vect{p}_{e}$ and $W$ is the work function of graphene. Within the dipole approximation,
\begin{align}\label{ARPESLM}
M_{f,i}\propto \bra{\mathrm{final}} \op{A}\cdot\op{p} \ket{\vect{k},i}, 
\end{align}
where $\op{A}$ is the vector potential of the incident photon, $\op{p}$ is the momentum operator, $\ket{\mathrm{final}}$ stands for the final state of the photoelectron and $\ket{\vect{k},i}$ denotes the wave function of the electron in the crystal. The latter is a linear combination of the sublattice Bloch states, Supplementary Equation \eqref{eqn:Bloch}, corresponding to wave vectors connected by a superlattice reciprocal vector $\vect{g}=\vect{G}'-\vect{G}$,
\begin{align}
\ket{\vect{k},i}=\sum_{\vect{g}}\sum_{l,X}c_{X,l}^{\vect{g},i}(\vect{k})\ket{\op{R}_{\theta_{l}}(\vect{k}+\vect{g}),X}_{l},
\end{align}
with the coefficients $c_{X,l}^{\vect{g},i}(\vect{k})$ provided by diagonalization of the Hamiltonian $\op{H}$, Supplementary Equation \eqref{ttlg}. Here, $\vect{g}$ is the moir\'{e} reciprocal superlattice vector. We approximate the final state with a plane wave (justified for incident photon energies above 50 eV \cite{gierz_nanolett_2012}) with momentum $\vect{p}_{e}=(\vect{p}_{e}^{\parallel},p_{e}^{\perp})$, so that
\begin{align*}
M_{f,i}\propto \sum_{\vect{g}}\sum_{l,X}c_{X,l}^{\vect{g},i}(\vect{k})\bra{e^{\tfrac{i}{\hbar}\vect{p}_{e}^{\parallel}\cdot\vect{r}}e^{\tfrac{i}{\hbar}p_{e}^{\perp}z}} \op{A}\cdot\op{p} \ket{\op{R}_{\theta_{l}}(\vect{k}+\vect{g}),X}_{l}.
\end{align*} 

Following from the chiral properties of the graphene wave function, the light-matter interaction $\op{A} \cdot \op{p}$ leads to an angle-dependent phase difference between the distinct atomic orbitals in the matrix element, $e^{i\varphi_{X,l}}$ \cite{liu_prl_2011, gierz_prb_2011, hwang_prb_2011} (note that this includes the effect of changing the sign of $\gamma_{1}$ \cite{hwang_prb_2011}). Hence,
\begin{align*}\begin{split}
M_{f,i} & \propto \sum_{\vect{g}}\sum_{l,X} e^{i\varphi_{X,l}} c_{X,l}^{\vect{g},i}(\vect{k})\braket{e^{\tfrac{i}{\hbar}\vect{p}_{e}^{\parallel}\cdot\vect{r}}e^{\tfrac{i}{\hbar}p_{e}^{\perp}z}|\op{R}_{\theta_{l}}(\vect{k}+\vect{g}),X}_{l} \\
& = \sum_{\vect{g}}\sum_{l,X, \vect{G}_l} e^{i\varphi_{X,l}}c_{X,l}^{\vect{g},i}(\op{R}_{-\theta_l}(\vect{p}_e^\parallel/\hbar+\vect{G}_l) - \vect{g}) e^{i \vect{G}_l \cdot \vect{\tau}_{X,l}}e^{-\tfrac{i}{\hbar}p_{e}^{\perp}z_l} \tilde{\phi}\left(\op{R}_{\theta_{l}}(\vect{k}+\vect{g})- \vect{G}_l, p_e^\perp/\hbar\right),
\end{split}\end{align*}
where
\begin{align*}
\tilde{\phi}\left(\vect{p}_{e}^{\parallel}/\hbar,{p}_{e}^{\perp}/\hbar\right) = \int d\vect{r}\, dz\, e^{-\tfrac{i}{\hbar} \vect{p}_{e}^{\parallel} \cdot \vect{r}} e^{-\tfrac{i}{\hbar} p_e^\perp z} \phi(\vect{r},z),
\end{align*}
is the Fourier transform of the $p_{z}$ orbital $\phi(\vect{r},z)$.

Due to the rotational symmetry of the $p_z$ orbital, $\tilde{\phi}\left(\vect{p}_{e}^{\parallel}/\hbar,{p}_{e}^{\perp}/\hbar\right)=\tilde{\phi}\left(|\vect{p}_{e}^{\parallel}/\hbar|,{p}_{e}^{\perp}/\hbar\right)$. Moreover, for the given photon energy, $\omega$, and work function, $W$, we have $p_e^\perp\gg|\vect{p}_e^\parallel|$ so that $\tilde{\phi}\left(\op{R}_{\theta_{l}}(\vect{k}+\vect{g})- \vect{G}_l, p_e^\perp/\hbar\right)$ can be approximated by a constant and dropped. Finally, in this work we only study points for which $\vect{G}_{l}=0$. As a result, 
\begin{align} \label{eq:Matrix4}
I \propto \sum_{i}\left|  \sum_{\vect{g}}\sum_{l,X}c_{X,l}^{\vect{g},i}(\op{R}_{-\theta_l}\vect{p}_e^\parallel - \vect{g}) e^{i\varphi_{X,l}} e^{-\tfrac{i}{\hbar}p_{e}^{\perp} l c_0}  \right|^2   \delta (\omega+\varepsilon_{i,\vect{k}}-W-\varepsilon_{\vect{p}_{e}}),
\end{align}
where we used the fact that $z_l = l c_0$. We combine both phases $\exp({i\varphi_{X,l}})$ and $\exp({-\tfrac{i}{\hbar}p_{e}^{\perp} l c_0})$ into a single factor,
\begin{align}
    e^{i\varphi_{X,l}} e^{-\tfrac{i}{\hbar}p_{e}^{\perp} l c_0} =e^{i\alpha_{X,l}}, 
\end{align}
which we fit to experiment. To note, the experimental data in this work suggests that the phase difference between the aligned graphene  layers is approximately $e^{i\pi}$.

Finally, we model the Dirac delta in Supplementary Equation \eqref{eq:Matrix4} with a Lorentzian
\begin{align}
\delta( \omega+\varepsilon_{i,\vect{k}}-W-\varepsilon_{\vect{p}_{e}})\rightarrow \dfrac{1}{\pi} \dfrac{\gamma}{( \omega+\varepsilon_{i,\vect{k}}-W-\varepsilon_{\vect{p}_{e}} )^2 - \gamma^2},   
\end{align}
with half-width-half-maximum $\gamma$ (the value of which, $\gamma\approx 0.17$ eV, we obtain by comparison to the experimental data).

%\newpage
%%\section{Simulating ARPES spectra of small-angle twisted bilayer graphene}
%\begin{flushleft}
%{\bf Supplementary Note 4: Simulating ARPES spectra of small-angle twisted bilayer graphene}
%\end{flushleft}
%
%In this section, we use our model, together with the parametrization of $t(\vect{r},z)$ as presented in the main text, to simulate ARPES spectra of twisted bilayer graphene with small twist angles in the vicinity of the magic angle $\theta_{c}=1.1^{\circ}$, recently measured experimentally \cite{lisi_arxiv_2020, utama_arxiv_2019}. We simulate ARPES intensity along the $k$-space cut connecting the Brillouin zone corners of the two layers, $\vect{K}_{2}$-$\vect{K}_{1}$ as introduced in the main text of our paper. We choose the spectral broadening as well as wave vector and energy ranges to match the experimental data (unfortunately, neither of the experimental works zoom in on the flat band features). The results, shown in Supplementary Figure 1 for $\theta=1.34^{\circ}$ \cite{lisi_arxiv_2020} in (a) and $\theta=0.96^{\circ}$ \cite{utama_arxiv_2019} in (b), are in qualitative agreement with the presented experimental data. We do not make comparisons to the constant energy maps presented in \cite{lisi_arxiv_2020} as it is unclear to us exactly what the symmetrization procedure is that is used to generate these maps.

\newpage
%\section{Fitting experimental ARPES data}
\begin{flushleft}
{\bf Supplementary Note 4: Fitting experimental ARPES data}
\end{flushleft}

In most cases, our graphene samples are electron-doped so that the Dirac and neutrality points are located below the Fermi level. To determine the energy position of the Dirac and neutrality points, we (i) select a point in the reciprocal space close to the anticipated location of the Brillouin zone corners, (ii) find the average between binding energies of intensity maxima corresponding to states in the valence and conduction bands. To check consistency, we repeat this procedure for several reciprocal space points. Similarly, we look at ARPES intensity along a cut in the reciprocal space passing through the assumed position of the Dirac/neutrality point and exploit circular symmetry of the low-energy electronic dispersion to triangulate their position. Once the locations of the Brillouin zone corners for the Bernal bilayer and twisted monolayer are established, the twist angle can be obtained by measuring the distance in reciprocal space between them (as explained in the main text, this distance is directly governed by the twist angle).   

As part of our data analysis, we extract the differences between on-site potentials on different graphene layers. For Bernal bilayer, we investigate the intensity at the corner of its Brillouin zone as a function of energy as the gap in the electronic dispersion is equal to $2|u|$ \cite{mccann_prl_2006}. The neutrality point is located halfway in the gap. In turn, the energy shift between the neutrality point and the position of the Dirac point of layer 3 defines $\Delta$.

Finally, in our theoretical analysis, we assume such an orientation of the graphene layers that one of the Brillouin zone corners of the first layer is located at $\vect{K}=\tfrac{4\pi}{3a}(1,0)$. In comparison, depending on the experimental geometry, the Brillouin zones of the graphene layers can be rotated arbitrarily about the $\Gamma$ point. Once the positions of the Dirac/neutrality points have been determined, we apply a global rotation about the $\Gamma$ point to map the positions of the neutrality points in the theoretical model onto the experimental features.

\newpage
\begin{flushleft}
{\bf Supplementary Figure 1: Simulating ARPES spectra of small-angle twisted bilayer graphene}
\end{flushleft}
\begin{figure*}[h]
    \centering
    \includegraphics[width = 1.0\linewidth]{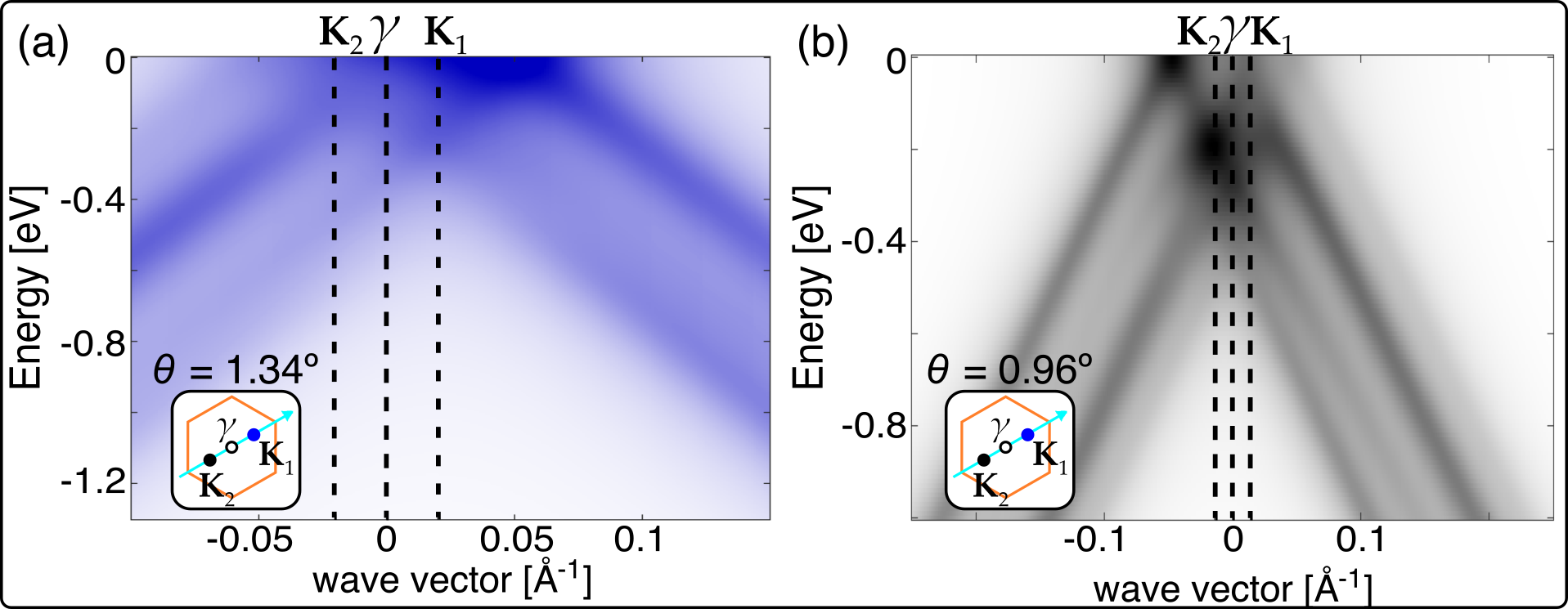}
    \caption{\textbf{ARPES of twisted bilayer graphene close to magic angle.} Simulation of ARPES intensity for twisted bilayer graphene with (a) $\theta=1.34^{\circ}$ and (b) $\theta=0.96^{\circ}$ along the $\vect{K}_{2}$-$\vect{K}_{1}$ cut as introduced in the main text and to be compared with experimental data shown in Ref.~\cite{lisi_arxiv_2020} and \cite{utama_arxiv_2019}, respectively. Above, we used our model with the parametrization of $t(\vect{r},z)$ as presented in the main text. We choose the spectral broadening as well as wave vector and energy ranges to match the experimental data (unfortunately, neither of the experimental works zoom in on the flat band features). The results are in qualitative agreement with the presented experimental photoemission intensity. We do not make comparisons to the constant energy maps presented in \cite{lisi_arxiv_2020} as it is unclear to us exactly what the symmetrization procedure is that is used to generate these maps.}
    \label{fig:S1}
\end{figure*}

\newpage
%\section{Comments on applicability of our approach to stacks of other 2D crystals}

%\begin{flushleft}
%{\bf Supplementary References}
%\end{flushleft}